\documentclass[preprint]{iucr}              

     \journalcode{A}              

\begin{document}                  



\title{Direct Phasing of Nanocrystal Diffraction}


\cauthor{Veit}{Elser}{ve10@cornell.edu}{address if different from \aff}

\aff{Department of Physics, Cornell University, Ithaca, NY 14853-2501 \country{USA}}









\maketitle                        

\begin{center}
Dedicated to the memory of David Sayre
\end{center}

\begin{synopsis}
We introduce a new direct phasing method that exploits small shifts in the positions of Bragg peaks in a nanocrystal diffraction pattern.
\end{synopsis}

\begin{abstract}
Recent experiments at free-electron laser x-ray sources have been able to resolve the intensity distributions about Bragg peaks in nanocrystals of large biomolecules. Information derived from small shifts in the peak positions augment the Bragg samples of the particle intensity with samples of its gradients. Working on the assumption that the nanocrystal is entirely generated by lattice translations of a particle, we develop an algorithm that reconstructs the particle from intensities and intensity gradients. Unlike traditional direct phasing methods that require very high resolution data in order to exploit sparsity of the electron density, our method imposes no constraints on the contrast other than positivity and works well at low resolution. We demonstrate successful reconstructions with simulated \textit{P1} lysozyme nanocrystal data down to a signal-to-noise ratio of 2 in the intensity gradients.
\end{abstract}


\section{Introduction}

To reconstruct the electron density within the unit cell of a crystal one needs to sum a Fourier series, each term associated with one Bragg peak of a diffraction pattern. This reconstruction process is severely under-constrained because the data provided by the Bragg intensities determine only the amplitudes and not the phases of the Fourier series coefficients. The so-called direct methods (Hauptman \& Karle, 1953) were developed to offset this data insufficiency. All of these methods impose additional constraints on the electron density, the most important of which, atomicity, asserts that at sufficient resolution the nonzero density regions are sparse in the unit cell. But because data at atomic resolution is rarely achieved with complex structures, the application of standard direct methods has been largely limited to small-molecule crystals.

An important new perspective on the crystallographic phase problem was brought to light by David Sayre (Sayre, 1952). His proposal applies to the case of molecular crystals, that is, where the motif in each unit cell has an integrity even in isolation. The continuous diffraction intensity of the motif is then well defined, and the Bragg peak intensities represent discrete samples of the continuous pattern. By invoking Shannon's sampling theorem, Sayre argued that while the Bragg samples are insufficient to reconstruct the continuous intensity, augmenting the samples to include half-integral lattice positions would be sufficient. In the case of a centrosymmetric motif, \textit{i.e.} a molecule with inversion symmetry, the phase of the Fourier transform is just a sign whose changes could then be easily determined from the surfaces where the continuous intensity vanishes.

In the decades following  Sayre's four-paragraph paper, his proposal has been developed in several ways. Firstly, Millane exploited the automatic existence of intensity samples between Bragg peak positions when the motif has non-crystallographic symmetry (Millane, 1993), as in the case of a virus crystal. Secondly, the reconstruction of phase from a continuous intensity, even in the non-centrosymmetric case, is a well posed problem for which there are now efficient solution methods (Thibault \& Elser, 2010). In this paper we propose yet another scheme for gaining access to the continuous intensity and thereby the molecular structure. The new method calls for samples of the \textit{gradients} of the continuous intensity, in addition to its values, at the Bragg peak positions. Intensity gradients can be inferred from small shifts in the peak positions when the crystal is small. The detection of such shifts is for the first time feasible with x-ray free electron lasers, where the high intensity of the source has produced diffraction signals from \textit{nanocrystals} comprising as few as $10^3$ unit cells (Chapman \textit{et al.}, 2011).

\section{Definition of a nanocrystal}

To keep the analysis simple, in this study we consider only nanocrystals whose macroscopic forms would have only one particle, typically one molecule, per unit cell. That is, our nanocrystals will have electron densities generated as translates of a particle density $\tilde{\rho}(\mathbf{r})$ by a subset $\mathcal{N}$ of vectors of some Bravais lattice:
\begin{equation}\label{nanoxtaldef}
\rho_\mathrm{nano}(\mathbf{r})=\sum_{\mathbf{n}\in \mathcal{N}}\tilde{\rho}(\mathbf{r}-\mathbf{n}).
\end{equation}
As most protein molecules have no symmetry, in the case of protein nanocrystals this definition restricts our analysis to proteins that crystallize in space group \textit{P1}. 

Proteins often crystallize with higher symmetry space groups by forming more symmetrical unit cell motifs from multiple molecules. It might seem that this case reduces to the single molecule case when the aggregate molecular motif is treated as a single particle. However, there is non-uniqueness in the definition of the particle, with each choice giving a different continuous diffraction intensity that the nanocrystal data is supposed to provide access to. Nanocrystals generated by translations of the non-uniquely defined particle are identical in the bulk, differing only on the surface. Thus not only does the multiple molecule case present new challenges, this case highlights a sensitivity to the nanocrystal surface structure that even the single molecule case is not immune to. We examine this issue next.

Figure 1 shows two electron density sections of lysozyme at low resolution. The image on the right shows a single molecule, as it would appear in isolation, while the image on the left shows parts of different molecules translated so as to lie within the same unit cell of a \textit{P1} crystal.
A physical nanocrystal, say of size $3\times 3\times 3$ unit cells, would be built from translates of a physical molecule and correspond to the image shown in Figure 2.
One could also construct a density by applying the same set of translations to the unit cell shown on the left in Figure 1; the result is shown in Figure 3. Clearly the physical (Fig. 2) and unphysical (Fig. 3) nanocrystals are indistinguishable in their interiors and differ only at their surfaces. Our analysis of nanocrystal diffraction patterns, based on the model expressed by equation (\ref{nanoxtaldef}), will be able to distinguish between the two unit cell motifs (physical and unphysical) because these have different continuous intensity patterns as shown in Figure 4.

The intensities (Fig. 4) of the two unit cell motifs (Fig. 1) agree at all the Bragg peak positions but deviate already in the gradients at these points. Our reconstruction method relies on an accurate determination of these gradients from nanocrystal diffraction. However, we have seen that the corresponding nanocrystals (Figs. 2 and 3) differ only at their surfaces. The converse side of this observation is that any systematic modification of the nanocrystal surface will compromise the successful application of our method. Such modifications might include significant rotations/translations or enhanced thermal motion of the surface layer molecules and cofactor binding at the surface. Since the surface equilibrium structure of protein nanocrystals is poorly understood at the present time, 
the viability of our method needs to be put to the test empirically, by extracting the intensity gradients as described in the next Section and comparing with the gradients of known structures.

\section{Extracting intensity gradients from diffraction data}

We will be working in the reciprocal space basis associated with the Miller indices of the macroscopic crystal form and with the corresponding dual or ``fractional" coordinates for positions in the unit cell. For example, the relationship between the diffraction amplitude at a Bragg peak and the electron density is expressed as
\begin{equation}\label{bragg}
F(\mathbf{q})=\int d\mathbf{r}\, \rho(\mathbf{r}) \exp{(2\pi i \mathbf{q}\cdot\mathbf{r})},
\end{equation}
where $\mathbf{q}$ is a vector of integers (the Miller indices) and the integration region of $\mathbf{r}$ (unit cell) is a unit cube. All our renderings of electron density will be distorted as a result of the fractional coordinate system. However, use of this system has no effect on our reconstruction method which makes no use of metrical structure in the electron density.

In order to extend the definition of the Bragg amplitudes (\ref{bragg}) to continuous $\mathbf{q}$ we define the \textit{particle unwrapping function} $\mathbf{u}(\mathbf{r})$ which relates the particle density $\tilde{\rho}$ in (\ref{nanoxtaldef}) to the unit-cell electron density function $\rho$ of (\ref{bragg}): 
\begin{equation}
\rho(\mathbf{r})=\tilde{\rho}(\mathbf{u}(\mathbf{r})).
\end{equation}
The unwrapping function $\mathbf{u}(\mathbf{r})$ in effect undoes the lattice translations that move every piece of the particle into a single unit cell. A comparison of these densities is shown in Figure 1.
We now replace equation (\ref{bragg}) with
\begin{eqnarray}
F(\mathbf{q})&=&\int d\mathbf{r}\, \tilde{\rho}(\mathbf{u}(\mathbf{r})) \exp{(2\pi i \mathbf{q}\cdot\mathbf{u}(\mathbf{r}))}\\
&=&\int d\mathbf{r}\, \rho(\mathbf{r}) \exp{(2\pi i \mathbf{q}\cdot\mathbf{u}(\mathbf{r}))},\label{offbragg}
\end{eqnarray}
and obtain an expression for the single particle diffraction amplitude valid at arbitrary $\mathbf{q}$. At integral $\mathbf{q}$ expressions (\ref{bragg}) and (\ref{offbragg}) agree because $\mathbf{u}(\mathbf{r})$ and $\mathbf{r}$ always differ by an integer vector. However, the gradient of the new amplitude
\begin{equation}
\nabla F(\mathbf{q})=2\pi i \int d\mathbf{r}\, \mathbf{u}(\mathbf{r})\rho(\mathbf{r}) \exp{(2\pi i \mathbf{q}\cdot\mathbf{u}(\mathbf{r}))},
\end{equation}
is changed by the unwrapping function even at integral $\mathbf{q}$. As we shall see below, the nanocrystal diffraction intensity provides access to the gradient  of the single particle intensity at the Bragg peak samples:
\begin{equation}
\nabla I(\mathbf{q})=\nabla |F(\mathbf{q})|^2=4\pi\, \mathrm{Im}(F(\mathbf{q})\mathbf{G}^\ast(\mathbf{q})),
\end{equation}
where we have introduced a new vector of amplitudes valid at integral $\mathbf{q}$:
\begin{equation}
\mathbf{G}(\mathbf{q})=\int d\mathbf{r}\, \mathbf{u}(\mathbf{r})\rho(\mathbf{r}) \exp{(2\pi i \mathbf{q}\cdot\mathbf{r})}.
\end{equation}

According to our definition (\ref{nanoxtaldef}), the diffraction intensity of the nanocrystal is
\begin{equation}
I_\mathrm{nano}(\mathbf{q})=|S(\mathbf{q})|^2\, I(\mathbf{q}),
\end{equation} 
where
\begin{equation}\label{nanostruc}
S(\mathbf{q})=\sum_{\mathbf{n}\in \mathcal{N}}\exp{(2\pi i \mathbf{q}\cdot\mathbf{n})}
\end{equation}
is the structure factor associated with the lattice translations that generate the nanocrystal. Whereas the details of the structure factor $S(\mathbf{q})$ depend on the set $\mathcal{N}$ that defines the nanocrystal, the following general characteristics apply to the structure factor of any nanocrystal of sufficient size: (i) It has the periodicity of the reciprocal lattice, and (ii) its magnitude is a symmetric function of the deviation $\Delta\mathbf{q}$ from the nearest Bragg vector and decays rapidly with the magnitude of $\Delta\mathbf{q}$. In free electron laser experiments (Chapman \textit{et al.}, 2011) the nanocrystal data will be collected from a very large ensemble of nanocrystals, all of which should differ only in the specific set $\mathcal{N}$. From the experiments we therefore obtain
\begin{equation}\label{nanodiff}
I_\mathrm{exp}(\mathbf{q})=s(\mathbf{q})\, I(\mathbf{q}),
\end{equation}
where
\begin{equation}
s(\mathbf{q})=\langle |S(\mathbf{q})|^2\rangle
\end{equation}
is the \textit{average crystal shape} function of the nanocrystal ensemble. Like each individual structure factor, the shape function is periodic and strongly peaked at the Bragg peak positions. Its width is of order the reciprocal nanocrystal diameter (number of unit cells).

Spence and coworkers have proposed using (\ref{nanodiff}) directly, that is, dividing out the average crystal shape function to obtain the particle intensity (Spence \textit{et al.}, 2011). However, the small width of the shape function limits the window to the particle intensity to small neighborhoods about the Bragg peak positions, and the straightforward application of this approach is problematic. Our method is to use the small width of the shape function as justification for the following approximation of the intensity about a Bragg peak $\mathbf{q}_0$:
\begin{equation}\label{peakfit}
I_\mathrm{exp}(\mathbf{q})\approx s(\mathbf{q})\left( I(\mathbf{q}_0)+(\mathbf{q}-\mathbf{q}_0)\cdot\nabla I(\mathbf{q}_0)\right).
\end{equation}
In a macroscopic crystal, where the width of $s(\mathbf{q})$ cannot be resolved, the product $s(\mathbf{q})(\mathbf{q}-\mathbf{q}_0)$ is effectively zero and the second term in (\ref{peakfit}) is negligible. The procedure then is to integrate $I_\mathrm{exp}(\mathbf{q})$ about $\mathbf{q}_0$ and thereby obtain the Bragg sample of the particle intensity $I(\mathbf{q}_0)$ (scaled by the  integral of the shape function). In the case of nanocrystals one can hope to go one step further by also including the gradient of the particle intensity in the fit to the data. 

The qualitative manifestation of the gradient term correction is to shift Bragg peaks in directions that increase the particle intensity. We can use this as the basis of a fitting scheme for extracting both the values and gradients of the particle intensity. The first task is to obtain a good estimate of the shape function $s(\mathbf{q})$ up to an irrelevant scale factor. This can be done by averaging several strong Bragg peaks after applying shifts to center them at a precise Bragg peak position, $\mathbf{q}_0$. Because the shape function is symmetric, the shifts are given as the average of $\mathbf{q}-\mathbf{q}_0$ weighted by intensity. Once a high quality shape function is obtained it is used to fit each Bragg peak intensity distribution according to equation (\ref{peakfit}). Assuming that the error in the fit is dominated by uncorrelated noise in $I_\mathrm{exp}(\mathbf{q})$ rather than errors in the shape function, a least squares procedure would be applied to samples $\mathbf{q}$ in the neighborhood of each peak. The result of this data reduction scheme is a list of particle intensity values and gradients at each of the Bragg peaks; an example is given in Table 1.

An intensity data set augmented by gradient information, as in Table 1, overcomes Sayre's sampling-theorem deficit (Sayre, 1952) by a generous amount. Recall that without the gradient information the reconstruction of the particle was under-constrained by a factor of two. Now, assuming the gradients can be determined with sufficient accuracy, the fourfold increase in the number of data implies the reconstruction is over-constrained by a factor of two. A reconstruction algorithm that uses the new data is described in the following two Sections.

Since so much depends on the accurate determination of the ensemble-averaged nanocrystal intensity $I_\mathrm{exp}(\mathbf{q})$, in particular its \textit{distribution} about Bragg peaks, we finish this Section on this topic. Single-shot free-electron laser data give Ewald-sphere slices of the intensity of individual nanocrystals. These have to be assigned an orientation relative to a fixed reciprocal lattice with a precision corresponding to the angular width of the Bragg peaks furthest from the origin. Likelihood-based methods, such as the EMC algorithm (Loh \& Elser, 2009), can achieve high quality orientation classification even when the Ewald-sphere slice has poor signal. However, in the classification of data from one nanocrystal the likelihood function should be derived from the specific structure factor of that same nanocrystal, which is not available. What can be done instead is to use the ensemble-averaged shape function $s(\mathbf{q})$ as a proxy for the nanocrystal-specific $|S(\mathbf{q})|^2$. Given the strong symmetry constraints on these two functions (same periodicity, centro-symmetry about Bragg positions) they should lead to almost identical likelihood functions on the orientation, especially when the data contain signal at multiple peaks. The nanocrystal-intensity-averaging and particle-intensity-gradient-extraction procedures can be validated with experiments on particles of known structure. However, the cause of an unsuccessful outcome might not be the data reduction procedure, as discussed above, but nanocrystal surface effects as described in the previous Section. 

\section{Four-replica reconstruction scheme}

In the previous Section we saw that nanocrystal diffraction data provides the following constraints on the particle Fourier transform $F$ and its vectorial supplement $\mathbf{G}$:
\begin{eqnarray}\label{dataconstraints}
|F(\mathbf{q})|&=&A(\mathbf{q})\label{dataconstraint1}\\
\mathrm{Im}\left(F(\mathbf{q})\mathbf{G}^\ast(\mathbf{q})\right)&=&A(\mathbf{q})\mathbf{B}(\mathbf{q}),\label{dataconstraint2}
\end{eqnarray}
where
\begin{eqnarray}
A(\mathbf{q})&=&\sqrt{I(\mathbf{q})}\\
\mathbf{B}(\mathbf{q})&=&\frac{1}{4\pi}\frac{\nabla I(\mathbf{q})}{\sqrt{I(\mathbf{q})}}.
\end{eqnarray}
The data constraints (\ref{dataconstraint1},\ref{dataconstraint2}) at different $\mathbf{q}$ are independent and are easily satisfied if the four complex numbers $F$ and $\mathbf{G}$ at each $\mathbf{q}$ can be treated as independent. We can achieve this by defining in addition to the density transform
\begin{equation}
F(\mathbf{q})=\int d\mathbf{r}\, \rho(\mathbf{r}) \exp{(2\pi i \mathbf{q}\cdot\mathbf{r})}
\end{equation}
the transform
\begin{equation}
\mathbf{G}(\mathbf{q})=\int d\mathbf{r}\, \mathbf{R}(\mathbf{r}) \exp{(2\pi i \mathbf{q}\cdot\mathbf{r})},
\end{equation}
and then imposing the constraint
\begin{equation}\label{consistency}
\mathbf{R}(\mathbf{r})=\mathbf{u}(\mathbf{r})\rho(\mathbf{r})
\end{equation}
at each $\mathbf{r}$ in the unit cell. Assuming for now the unwrapping function $\mathbf{u}(\mathbf{r})$ is known, equation (\ref{consistency}) is a linear constraint and also easy to satisfy.

The four functions $\rho$ and $\mathbf{R}$ are ``replicas" in the sense that from any one of them we essentially can infer any of the others. By granting independence to the replicas the data constraints that apply to their Fourier transforms are easily satisfied. Some sections of the unwrapping function and replicas for the lysozyme molecule are shown in Figure 5.

\subsection{Constraint projections}

To turn our reconstruction problem into an exercise in numerical constraint satisfaction we first have to define the space of variables. Suppose the available data (intensities and gradients) is contained in the Miller-index set
\begin{equation}\label{Q}
\mathcal{Q}=\{(h,k,l) \colon |h|\le N, |k|\le N, |l|\le N\}
\end{equation}
for some $N$.
We would then define the four complex-valued functions $F$ and $\mathbf{G}$ on four cubic grids of size $M=2N+1$; the discrete Fourier transforms of these (Friedel-symmetric) functions give us the four real-valued functions $\rho$ and $\mathbf{R}$ sampled on grids also of size $M$ in the unit cell. The two sets of functions are completely equivalent representations of the structure, with one set tailored for the data constraints (\ref{dataconstraint1},\ref{dataconstraint2}) and the other to the replica-consistency constraint (\ref{consistency}). Our reconstruction algorithm is based on projections and requires a definition of distance between one configuration of variables $(F, \mathbf{G})$ and another $(F', \mathbf{G}')$, or their Fourier counterparts. We will use the $L^2$ distance
\begin{eqnarray}\label{dist}
\Delta^2&=&\sum_{\mathbf{q}\in\mathcal{Q}}\;|F'(\mathbf{q})-F(\mathbf{q})|^2+\sigma^2 |\mathbf{G}'(\mathbf{q})-\mathbf{G}(\mathbf{q})|^2\\
&=&\frac{1}{M^3}\sum_{\mathbf{r}\in\mathcal{Q}/M}|\rho'(\mathbf{r})-\rho(\mathbf{r})|^2+\sigma^2 |\mathbf{R}'(\mathbf{r})-\mathbf{R}(\mathbf{r})|^2
\end{eqnarray}
which allows for adjusting the weighting of the gradient part of the configuration by the dimensionless metric scale parameter $\sigma$. This parameter can be removed from the definition of the distance by rescaling $\mathbf{G}$ and its Fourier transform $\mathbf{R}$, the result of this being a rescaling of the unwrapping function and the gradient data:
\begin{equation}
\mathbf{u}\to \sigma\mathbf{u}\qquad \mathbf{B}\to \sigma\mathbf{B}.
\end{equation}
All subsequent formulas will assume this rescaling has been made.

\subsubsection{Replica consistency projection}

There is one replica consistency constraint (\ref{consistency}) at each grid point $\mathbf{r}$ and it involves only the four replicas $(\rho(\mathbf{r}),\mathbf{R}(\mathbf{r}))$ at that point. We will therefore omit reference to $\mathbf{r}$ in the following. The projection $(\rho,\mathbf{R})\to (\rho',\mathbf{R}')$ to the constraint $\mathbf{R}'=\mathbf{u}\rho'$ is geometrically a projection to a line in a four dimensional space and is given by the stationary point of the distance to which a Lagrange multiplier term has been added:
\begin{equation}\label{dist1}
|\rho'-\rho|^2+|\mathbf{R}'-\mathbf{R}|^2+2\mathbf{m}\cdot(\mathbf{R}'-\mathbf{u}\rho').
\end{equation}
Stationarity of (\ref{dist1}) gives the pair of equations
\begin{eqnarray}
\rho'&=&\rho+\mathbf{m}\cdot\mathbf{u}\\
\mathbf{R}'&=&\mathbf{R}-\mathbf{m},\label{R'}
\end{eqnarray}
which together with the equation $\mathbf{R}'=\mathbf{u}\rho'$ allow us to eliminate all the primed variables to obtain the following equation for the Lagrange multiplier:
\begin{equation}\label{mequation}
\mathbf{m}+(\mathbf{m}\cdot\mathbf{u})\mathbf{u}=\mathbf{R}-\mathbf{u}\rho.
\end{equation}
Solving for $\mathbf{m}$ in equation (\ref{mequation}),
\begin{equation}
\mathbf{m}=\mathbf{R}-\left(\frac{\rho+\mathbf{u}\cdot\mathbf{R}}{1+\mathbf{u}\cdot\mathbf{u}}\right)\mathbf{u}
\end{equation}
and comparing this solution with (\ref{R'}) we obtain
\begin{equation}
\mathbf{R}'=\left(\frac{\rho+\mathbf{u}\cdot\mathbf{R}}{1+\mathbf{u}\cdot\mathbf{u}}\right)\mathbf{u},
\end{equation}
the coefficient of $\mathbf{u}$ being the projected $\rho'$. 

It is easy to modify this projection so that it imposes positivity of the electron density as well. In this case the constraint set is geometrically the half-line where $\rho'$ is positive. If the projection computed above gives a point with negative $\rho'$, then the nearest point of the true constraint set should be the endpoint of the half-line, or $\rho'=\mathbf{R}'=0$.

\subsubsection{Projection to the data constraints}

In analogy with the replica consistency constraints, the data constraints (\ref{dataconstraint1},\ref{dataconstraint2}) apply, independently for each $\mathbf{q}$, to the four complex numbers $(F(\mathbf{q}),\mathbf{G}(\mathbf{q}))$. We omit reference to $\mathbf{q}$ in the following.

Unlike the replica consistency constraints, the data constraints are nonlinear and projecting to the non-convex set requires more work. We begin our calculation of the projection $(F,\mathbf{G})\to(F',\mathbf{G}')$ by fixing $F$ and $F'$ and finding a stationary point with respect to $\mathbf{G}'$ of the Lagrange-multiplier augmented distance
\begin{equation}
|\mathbf{G}'-\mathbf{G}|^2+2\mathbf{n}\cdot\left(\mathrm{Im}(F'\mathbf{G}'^\ast)-A\mathbf{B}\right),
\end{equation}
where $\mathbf{n}$ is the real-valued multiplier. The value of $\mathbf{G}'$ at the stationary point,
\begin{equation}\label{G'}
\mathbf{G}'=\mathbf{G}+i F'\mathbf{n},
\end{equation}
when substituted into the constraint equation
\begin{equation}
\mathrm{Im}(F'\mathbf{G}'^\ast)=A\mathbf{B}
\end{equation}
gives the following solution for the Lagrange multiplier:
\begin{equation}\label{n}
\mathbf{n}=\mathrm{Im}(F'\mathbf{G}^\ast)/A^2-\mathbf{B}/A.
\end{equation}
We note that our solution (\ref{G'}) for $\mathbf{G}'$ is unique because the constraint on this variable is linear. 

The next step is to minimize the distance
\begin{equation}\label{dist2}
d(\phi)=|F'-F|^2+|\mathbf{G}'-\mathbf{G}|^2=|F'-F|^2+|\mathrm{Im}(F'\mathbf{G}^\ast)/A-\mathbf{B}|^2,
\end{equation}
 with respect to the phase angle of $F'=A\exp{(i\phi)}$. The second term in $d(\phi)$ was expressed in terms of $F'$ by using (\ref{G'},\ref{n}) and the constraint $|F'|=A$. Writing the two terms of the distance explicitly as a function of $\phi$ we obtain
 \begin{eqnarray}
 |F'-F|^2&=&-A F \exp{(-i\phi)}+\mathrm{c.c.}\quad+\cdots\\
 |\mathbf{G}'-\mathbf{G}|^2&=&\left(-i\,\mathbf{B}\cdot\mathbf{G} \exp{(-i\phi)}-\frac{1}{4}\mathbf{G}\cdot\mathbf{G} \exp{(-i2\phi)}\right)+\mathrm{c.c.}\quad+\cdots,
 \end{eqnarray}
 where c.c. denotes the complex-conjugate of the preceding expression and $\;\cdots\;$ represents $\phi$-independent terms. Defining the constants
 \begin{eqnarray}
 a\exp{(i\alpha)}&=&A F+i\,\mathbf{B}\cdot\mathbf{G}\label{item2a}\\
 b\exp{(i\beta)}&=&\frac{1}{4}\mathbf{G}\cdot\mathbf{G},\label{item2b}
 \end{eqnarray}
 the distance takes the compact form
 \begin{equation}
d(\phi)=-a\cos{(\phi-\alpha)}-b\cos{(2\phi-\beta)}+\quad\cdots\quad. 
 \end{equation}
Since the constraint on $F'$ is nonlinear, this distance can have more than one stationary point, depending on the ratio $b/a$ and the angles $\alpha$ and $\beta$. At the core of this projection we thus require the function
\begin{equation}\label{psidef}
\psi(c,\gamma)=\mathop{\mathrm{argmin}}_{\psi'} \left(-\cos{(\psi')}-c\cos{(2\psi'-2\gamma)}\right)
\end{equation}
of two real arguments. The argument $c$ is nonnegative and it is enough to consider $|\gamma|\le\pi/2$ since $\psi$ has periodicity $\pi$ in this argument. Figure 6 shows a plot of $\psi(c,\gamma)$. Limiting values are $\psi(0,\gamma)=0$ and $\psi(\infty,\gamma)=\gamma$; there is a discontinuity at $\gamma=\pm\pi/2$ for $c>1/4$.

Below is a summary of the projection to the data constraint:
\begin{enumerate}
\item Apply discrete Fourier transforms to each of the four replicas $(\rho,\mathbf{R})$ to obtain $(F,\mathbf{G})$.

\item Determine $a$, $\alpha$, $b$, and $\beta$ from equations (\ref{item2a},\ref{item2b}).

\item Obtain the phase of $F'$ as $\phi=\alpha+\psi(b/a,\beta/2-\alpha)$.

\item Evaluate the Lagrange multiplier $\mathbf{n}$ with equation (\ref{n}) and substitute into equation (\ref{G'}) to obtain $\mathbf{G}'$.

\item Apply the inverse Fourier transform to $(F',\mathbf{G}')$ to obtain $(\rho',\mathbf{R}')$.

\end{enumerate}
Steps 2-4 are applied at each $\mathbf{q}$ for which there is data. In the event that the data at some $\mathbf{q}$ is missing or unreliable, the corresponding amplitudes $(F,\mathbf{G})$ are left unchanged.
 
\section{Iterative reconstruction algorithm}

Our reconstruction algorithm uses the \textit{alternating directions method of multipliers} (ADMM) iteration scheme (Boyd \textit{et al.}, 2011) for the two constraint projections described in the preceding Section. Combining the four replicas into a single symbol, $\mbox{\boldmath$\rho$}=(\rho,\mathbf{R})$, the ADMM scheme is to cyclically iterate the following:
\begin{eqnarray}
\mbox{\boldmath$\rho$}_1&=&P_1(\mbox{\boldmath$\rho$}_2+\alpha\, \mathbf{z})\\
\mbox{\boldmath$\rho$}'_2&=&P_2(\mbox{\boldmath$\rho$}_1-\alpha\, \mathbf{z})\\
\mathbf{z}'&=&\mathbf{z}+\mbox{\boldmath$\rho$}'_2-\mbox{\boldmath$\rho$}_1.
\end{eqnarray}
The two constraint projections are denoted $P_1$ and $P_2$, one of which implements the replica consistency constraint while the other projects to the data constraint. We have chosen $P_1$ to be the replica consistency projection, although the other choice (which gives an inequivalent algorithm) was also found to perform well. The iteration rule updates three sets of variables: $\mbox{\boldmath$\rho$}_1$, $\mbox{\boldmath$\rho$}_2$ and $\mathbf{z}$. The first two, by construction, satisfy constraints 1 and 2 respectively. When we have $\mbox{\boldmath$\rho$}_1=\mbox{\boldmath$\rho$}'_2$ both constraints are satisfied, there is no change in $\mathbf{z}$, and the iteration arrives at a fixed point. During the search for the fixed point $\mathbf{z}$ grows in proportion to the current constraint incompatibility and provides a mechanism for the iterations to escape near-solutions of the type where the distance between the two constraint sets is a local minimum (but nonzero). The strength of the incompatibility on the next round of projections is controlled by the dimensionless parameter $\alpha$. It is clear that with $\alpha$ set to zero the ADMM scheme reduces to the simple alternation of projections. In the appendix we show that $\alpha=1$ is equivalent to the Douglas-Rachford algorithm, which is the form taken by the Fienup (or difference-map) iteration with $\beta=1$ (Fienup, 1982; Elser, 2003). In all the reconstructions presented below we have used $\alpha=0.5$. Every reconstruction was initialized with $\mathbf{z}=0$ and $\mbox{\boldmath$\rho$}_2$ set to a random set of replicas that have been projected to the data constraint. We monitor progress through the distance $\Delta=\|\mbox{\boldmath$\rho$}'_2-\mbox{\boldmath$\rho$}_1\|$ given by equation (\ref{dist}).

\subsection{Particle unwrapping}

Reconstructions fall into two classes, depending on whether the particle unwrapping function $\mathbf{u}(\mathbf{r})$ is known or unknown. When the particle is compact and the fraction of empty or solvent-occupied space is large, it is usually safe to assume the particle fits inside the unit cell parallelepiped without ``wrapping around". In that case we may use the trivial unwrapping function
\begin{equation}\label{trivialu}
\mathbf{u}(\mathbf{r})=\sigma \mathbf{r},\qquad \mathbf{r}\in\mathcal{Q}/M.
\end{equation}
However, in low solvent fraction crystals it will often be the case that the particle cannot be centered without wrapping around the parallelepiped. Moreover, in these cases there will be close interparticle contacts making the definition of the particle perimeter ambiguous even when the electron density is well reconstructed. The \textit{P1} crystal form of lysozyme (Wang \textit{et al.}, 2007) used in our simulations, with 27\% solvent fraction, is a good example of this situation. The particle unwrapping method described in this Section was developed over the course of simulations with this molecule but is completely general.

 Our unwrapping method begins with the trivial function (\ref{trivialu}) and refines it adiabatically based on averages of the reconstructed electron density. The update rule for $\mathbf{u}(\mathbf{r})$ is based on a simple representation by ``wrapping surfaces". There are three wrapping surfaces, one associated with each coordinate of the unit cell: $X(y,z)$, $Y(x,z)$, $Z(x,y)$. Each surface specifies where the corresponding component of the wrapping function changes discontinuously by $\pm \sigma$. Examples for the case of lysozyme are shown in the top panels of Figure 5. To make a precise definition we use the shifted modulus operation $\mathrm{mod}(1,X)$, where the statement
 \begin{equation}
 x'=x\;\mathrm{mod}(1,X),
 \end{equation}
 means that $x'$ is the unique translate of $x$ by an integer that lies in the half-open interval $(X-1,X]$. The $x$-component of the unwrapping function is defined as
 \begin{equation}\label{wsurface}
 u_x(x,y,z)=\sigma\left(x\;\mathrm{mod}(1,X(y,z))\right),
 \end{equation}
 and analogously for the other two components. For the trivial unwrapping function (\ref{trivialu}) all three surfaces are flat and have values $X=Y=Z=1/2$.

Our representation of the unwrapping function is not the most general possible, but is probably general enough for most applications. It is limited by having only one discontinuity on any line of length 1 parallel to one of the unit cell axes. For example, the single discontinuity of $u_x(x,y,z)$ as $x$ ranges from $-1/2$ to $+1/2$ (with $y$ and $z$ fixed) asserts that the identity of the particle changes exactly once per unit cell when this line is extended indefinitely in the crystal. An example of a crystal structure requiring a more complex unwrapping function is shown in Figure 7. The S-shaped particle in this structure packs in such a way that its unwrapping function has one component with three discontinuities per unit cell.

To specify the update rule for the unwrapping function we can focus on the update of the wrapping surfaces because these define the unwrapping function through (\ref{wsurface}). We refine the surfaces by a heuristic that has some similarities with the shrink-wrap procedure (Marchesini \textit{et al.}, 2003) of refining particle support in the non-crystalline case. However, we note that modifying the wrapping surfaces does not change the volume available to the particle, that is, add or remove constraints in the reconstruction. The wrapping surfaces are additional degrees of freedom that must be reconstructed from the available data (subject to ambiguity where the density vanishes). We reconstruct these surfaces, in analogy with the shrink-wrap method, by imposing a lower limit on the scale of their variation and refining the variations by the particle contrast that passes closest to the surfaces.

Our heuristic rule for moving the wrapping surface by $\Delta X(y,z)$ is based on the asymmetry between the density $\rho(X^+,y,z)$ just ``above" the surface and the density $\rho(X^-,y,z)$ just ``below" it. In the extreme case, where the density on one side of the surface is exactly zero, the surface can be moved by one grid spacing without violating any new constraints because none of the replicas are thereby changed. On the other hand, if the density on the other side of the surface is positive, then moving the surface and thereby increasing the volume available at the particle boundary can only help in satisfying constraints. Both the direction for moving the surface and the strength of the asymmetry is captured by a local force computed as follows:
\begin{equation}
f_x(y,z)=\langle\; \rho(X^-(y,z),y,z)-\rho(X^+(y,z),y,z)\;\rangle
\end{equation}
The superscripts $-$ and $+$ denote grid points on either side of $X$ and the angle brackets represent a block average of several ADMM iterations. After each block average, a Gaussian low-pass filter is applied to $f_x(y,z)$ and $X(y,z)$ is incremented by a positive multiple of the smoothed force function. The other two surfaces are updated analogously. 

\section{Reconstruction experiments with simulated \textit{P1} lysozyme data}

We tested our direct phasing method with simulated data derived from the \textit{P1} lysozyme structure of Wang \textit{et al.} (2007), a low solvent fraction crystal comprising 1001 non-hydrogen atoms in a single, 129-residue protein molecule. The (macro-crystal) Bragg intensities for this structure have been measured to very high resolution (0.65\AA, 187165 unique reflections), making this also a rare instance where atomicity-based direct methods can be successfully applied to a protein (Deacon \textit{et al.}, 1998). In our nanocrystal direct phasing simulation, by contrast, we truncated the data to a much lower resolution to reflect the as yet unknown reliability of intensity-gradient extraction from nanocrystal data. We used Miller index cutoff $N=17$ for a data set comprising $21437\times 4$ unique intensities and gradients. Moreover, a Gaussian low-pass filter was applied to the data so as not to introduce negative electron density as a result of series truncation. The actual resolution of the simulated data is therefore quite low, about {2\AA} as measured by the fading of the filtered intensities shown in Figure 8. The electron density reconstructed with the known phases at this resolution is shown in Figure 9 and is sufficient to identify the secondary structure elements of the protein.

Because errors in the extraction of the gradients will be the dominant source of experimental uncertainty, for simplicity we introduced noise only via the reduced gradient data $\mathbf{B}(\mathbf{q})$. We used uncorrelated additive Gaussian noise, uniform in $\mathbf{q}$; the signal-to-noise ratios (SNR) we quote are root-mean-square measures applied to $\mathbf{B}$ and the noise that was added. Also for the sake of simplicity the metric scale parameter was fixed at $\sigma=3$ in all the experiments. The quality of reconstructions at high noise might have been improved with a smaller $\sigma$, which controls the relative weight of the gradient constraints, that is,  constraints on $\mathbf{G}$ relative to $F$. A uniform density particle whose shape is a unit-diameter sphere in the fractional coordinate system will have equal power in $F$ and each component of $\mathbf{G}$ when $\sigma=\sqrt{20}$.  

Whereas our simulations made use of all the $\mathbf{q}\ne0$ data in our truncated set, in practice one might want to ignore those gradient data with uncertainty above some threshold. In such cases one could still use the Bragg intensity data $A(\mathbf{q})$ as a single constraint on $F(\mathbf{q})$, leaving $\mathbf{G}(\mathbf{q})$ unchanged.

All simulations were performed on a laptop computer running software written in Mathematica 7.0.

\subsection{Known wrapping}

When given the unwrapping function from the known protein envelope, the reconstruction algorithm was able to determine the phases quickly, apparently without having to do any exploration. The iteration series of the ADMM updates $\Delta$, shown in Figure 10, is almost independent of the random initial density. A steady state is reached after about 50 iterations, whereupon $\Delta$ fluctuates about a value that is set by the SNR. After 100 iterations we calculated the standard phase figure-of-merit (FOM)
\begin{equation}
\langle\cos{\Delta\phi}\rangle=\frac{\sum_{\mathbf{q}\ne 0} I(\mathbf{q}) \cos{\Delta\phi(\mathbf{q})}}{\sum_{\mathbf{q}\ne 0} I(\mathbf{q})},
\end{equation}
where $\Delta\phi(\mathbf{q})$ is the difference between the true and reconstructed phase, and the result is maximized with respect to a relative translation. Apparently the unwrapping function was sufficiently enantiomer specific that the only variability encountered in the reconstructions was a small translation. Table 2 gives FOM values as a function of SNR for three experiments at each noise value. The reconstructed electron density contours for $\mathrm{SNR}>2$ are essentially indistinguishable from the noise-free contours in the top panel of Figure 9. The lower panel of the same Figure shows a reconstruction with $\mathrm{SNR}=2$ and some deterioration at the level of secondary structure features.

\subsection{Unknown wrapping}

The reconstruction of a tightly packed particle, such as \textit{P1} lysozyme, is much more difficult when the unwrapping function is unknown. We have used the force-heuristic described above, starting with three flat wrapping surfaces and updating them every ten iterations of the ADMM density reconstruction. An example of the evolution of the surfaces in one reconstruction is shown in Figure 11. Many modifications of the surfaces are explored before the algorithm discovers the correct ones. In the case of lysozyme the surfaces define the enantiomer; the handedness of the reconstruction is randomly determined by the initial electron density.

Plots of the replica update magnitudes $\Delta$ for SNR 20, 10 and 5 are shown in Figure 12. The sudden drop in $\Delta$, coinciding with the discovery of the wrapping surfaces, is stochastic and in one case required 180 updates of the surfaces using our heuristic. Another difference, relative to the known unwrapping function case, is the higher fluctuating level of $\Delta$ in the post-solution-discovery steady state. We believe this is due to the smoothness of our wrapping surfaces, as imposed by filtering the force; the true surfaces have sharper features. Reducing the smoothness of the surfaces should reduce $\Delta$ in the phased particle and extend the reach of our method to higher noise. However, this comes at the expense of enlarging the search-space of wrapping surfaces. In Figure 13 we compare low resolution contours of a SNR 5 reconstruction with the true density. Phase FOM values are given in Table 3; as expected they are somewhat lower than the corresponding values for the known unwrapping function case.

\section{Conclusions}

Putting aside for the moment the practicality and reliability of experimental methods for extracting intensity gradient data from nanocrystal diffraction, the reconstruction method we have developed (i) is directly based on the available constraints, (ii) can be efficiently implemented by an iterative algorithm, and (iii) has been demonstrated in simulations even in the presence of significant noise. There is thus no theoretical obstacle in attempting a proof-of-principle experiment, say with lysozyme where the single-particle model used in this work is most likely applicable.

The main challenge for the first experiments will be to achieve high fidelity in the 3D assembly of the nanocrystal-ensemble intensity from noisy and non-oriented 2D sections. There has been much recent progress on this problem (Kirian \textit{et al.}, 2010). However, for the direct phasing work the demands on the quality of the 3D intensity will be much greater than what is required for extracting the integrated Bragg intensities. Still, it should be possible to validate procedures for extracting intensity gradients from peak shifts by working with known structures.

A greater challenge for the future is generalizing the reconstruction algorithm to allow for multiple particles per unit cell. The existence of multiple particles provides a natural mechanism for the nanocrystal to assume a surface structure that goes beyond the model used in the present work. An example of the phenomenon is shown in Figure 14. Here two copies of a non-symmetric particle (triangle) are arranged in the unit cell so as to give the crystal a $\pi$-rotation symmetry. However, not both copies need be present at the surface and depending on the crystal facet just one or the other might be the preferred form. This choice is a new degree of freedom that an extension of the present work would have to address. 


\appendix
\section{}
\subsection{Equivalence of iteration schemes}

To see the equivalence of the ADMM iteration with $\alpha=1$ and the $\beta=1$ difference-map (Fienup's hybrid input-output) we group the three updates as follows:
\begin{eqnarray}
\mbox{\boldmath$\rho$}'_2&=&P_2(\mbox{\boldmath$\rho$}_1- \mathbf{z})\\
\mathbf{z}'&=&\mathbf{z}+\mbox{\boldmath$\rho$}'_2-\mbox{\boldmath$\rho$}_1\\
\mbox{\boldmath$\rho$}'_1&=&P_1(\mbox{\boldmath$\rho$}'_2+\mathbf{z}').
\end{eqnarray}
To start the iteration with this grouping we would initialize $\mbox{\boldmath$\rho$}_1$ and $\mathbf{z}$ and cycle through the equations in the order written. Now define the variable $\mathbf{x}=\mbox{\boldmath$\rho$}_1-\mathbf{z}$, which we would initialize accordingly, and rewrite the three steps of the ADMM iteration with $\mathbf{x}$ replacing $\mathbf{z}$:
\begin{eqnarray}
\mbox{\boldmath$\rho$}'_2&=&P_2(\mathbf{x})\label{eq1}\\
\mbox{\boldmath$\rho$}'_1-\mathbf{x}'&=&\mbox{\boldmath$\rho$}_1-\mathbf{x}+\mbox{\boldmath$\rho$}'_2-\mbox{\boldmath$\rho$}_1=\mbox{\boldmath$\rho$}'_2-\mathbf{x}\label{eq2}\\
\mbox{\boldmath$\rho$}'_1&=&P_1(\mbox{\boldmath$\rho$}'_2+\mbox{\boldmath$\rho$}'_1-\mathbf{x}')=P_1(2\mbox{\boldmath$\rho$}'_2-\mathbf{x}).\label{eq3}
\end{eqnarray}
We simplified the argument of the projection in (\ref{eq3}) by using (\ref{eq2}). If we now rewrite (\ref{eq2}) while substituting the expressions for $\mbox{\boldmath$\rho$}'_1$ and $\mbox{\boldmath$\rho$}'_2$ from the other two equations we see that we get an update rule that only makes reference to $\mathbf{x}$:
\begin{equation}
\mathbf{x}'=\mathbf{x}+P_1(2 P_2(\mathbf{x})-\mathbf{x})-P_2(\mathbf{x}).
\end{equation}
This is the difference-map update with $\beta=1$. 
 

\newpage

\ack{Acknowledgements}

This work was begun in 2009 while I was a sabbatical visitor at the Center for Free Electron Laser Science at DESY. My remarks in the conclusions Section are largely the result of discussions with the CFEL team four years later. I thank in advance the referees for their help with citations. This work was supported by DOE Grant DE-FG02-11ER16210.



     
\newpage

\begin{table}
\caption{Some intensities and intensity gradients of the lysozyme molecule.}
\begin{center}
\begin{tabular}{rrr|rrrr}      
 $h$    & $k$        & $l$       & $I$ & $\partial_h I$   & $\partial_k I$& $\partial_l I$  \\
\hline
 0&0&0      & 100.00      & 0.00  & 0.00 & 0.00      \\
 1&0&0      & 2.51      & -25.64  & -4.83 & -8.30      \\
 0&1&0      & 0.41      & -0.18  & -11.48 & 4.69      \\
 0&0&1      & 1.38      & -4.89  & 6.15 & -19.14      \\
 
 1&1&0      & 0.13      & -1.37  & -0.89 & -1.10      \\
 1&-1&0      & 0.21      & -1.77  & 0.67 & -1.03      \\
 
 1&0&1      & 0.22      & -0.56  & -0.21 & -0.17      \\
 1&0&-1      & 0.08      & 2.10  & 1.43 & -2.15      \\
 
 0&1&1      & 0.01      & 0.35  & 0.11 & 0.46      \\
 0&1&-1      & 1.38      & 1.27  & -0.62 & 2.62      \\
\end{tabular}
\end{center}
\end{table}

\begin{table}
\caption{Phase figure-of-merit (FOM) as a function of SNR in the gradient data, in lysozyme reconstructions (three trials) with known unwrapping function.}
\begin{center}
\begin{tabular}{r|l}      
 SNR    & $\langle\cos{\Delta\phi}\rangle$\\
\hline
 20& 0.99,  0.99,  0.99\\
 10& 0.98, 0.98, 0.98\\
 5& 0.95, 0.95, 0.95\\
 2& 0.83, 0.85, 0.86\\
 1& 0.40, 0.50, 0.55
\end{tabular}
\end{center}
\end{table}

\begin{table}
\caption{Same as Table 2 but for reconstructions with unknown unwrapping function.}
\begin{center}
\begin{tabular}{r|l}      
 SNR    & $\langle\cos{\Delta\phi}\rangle$\\
\hline
 20& 0.98, 0.97, 0.98 \\
 10& 0.97, 0.96, 0.96\\
 5& 0.93, 0.95, 0.93\\
\end{tabular}
\end{center}
\end{table}


\newpage

\begin{figure}
\caption{Two electron density maps of lysozyme. \textit{Right}: the single molecule density $\tilde{\rho}$ and its relation to the unit cell of the crystal; \textit{left}: parts of different molecules translated to lie within one unit cell together comprise the density $\rho$.}
\includegraphics{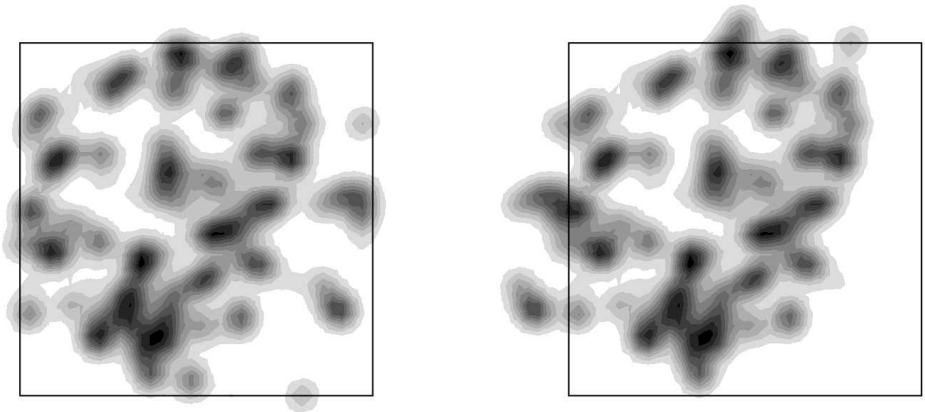}
\end{figure}

\begin{figure}
\caption{Section of a $3\times 3\times 3$ nanocrystal formed from translates of a physical molecule (Fig. 1 \textit{right}).}
\includegraphics{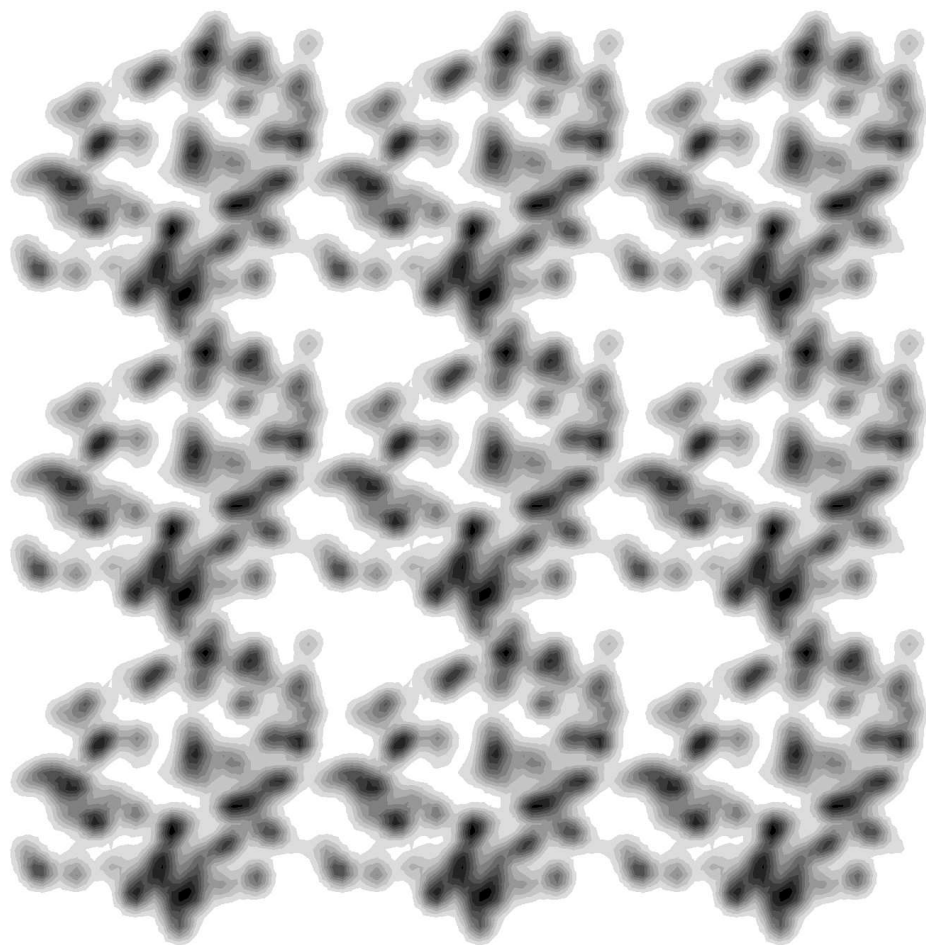}
\end{figure}

\begin{figure}
\caption{Same as Figure 2 but for translates of an unphysical unit cell motif (Fig. 1 \textit{left}).}
\includegraphics{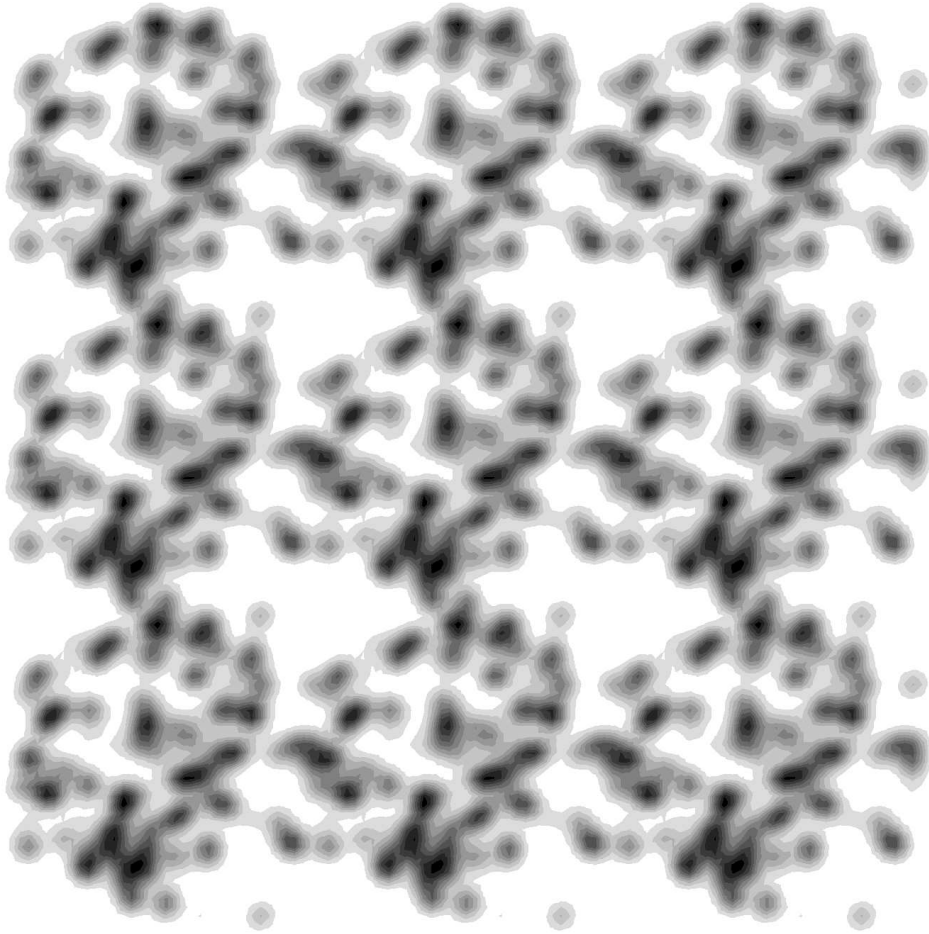}
\end{figure}

\begin{figure}
\caption{Continuous intensities corresponding to the two unit cell motifs of Figure 1. The white points give the locations of the Bragg peaks.}
\includegraphics{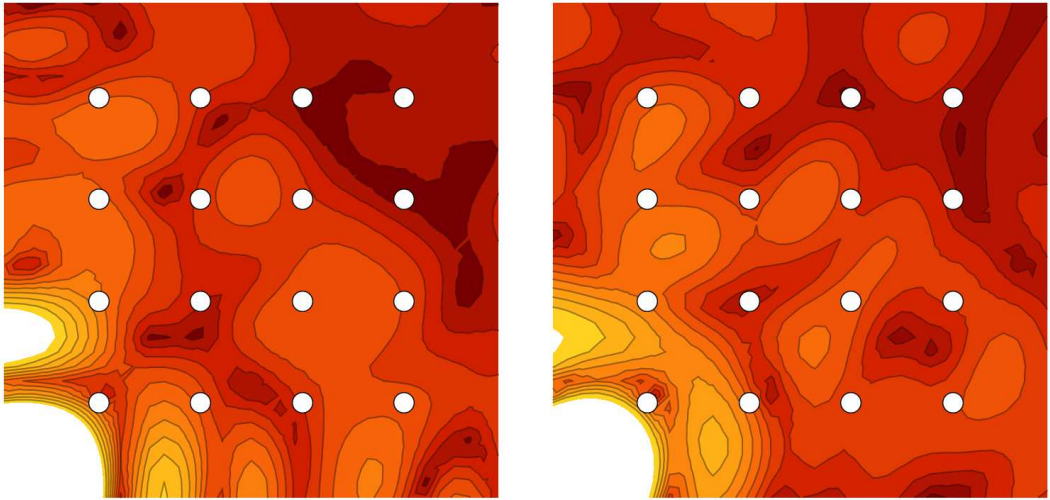}
\end{figure}

\begin{figure}
\caption{\textit{Top panels}: $x$ and $y$ components of the unwrapping function $\mathbf{u}$ for the lysozyme molecule. The top left panel shows $u_x$ is negative at the right side of the unit cell because the molecule wraps around the cell. \textit{Lower panels}: $x$ and $y$ components of $\mathbf{R}=\mathbf{u}\rho$ (two ``replicas" of the lysozyme density).}
\includegraphics{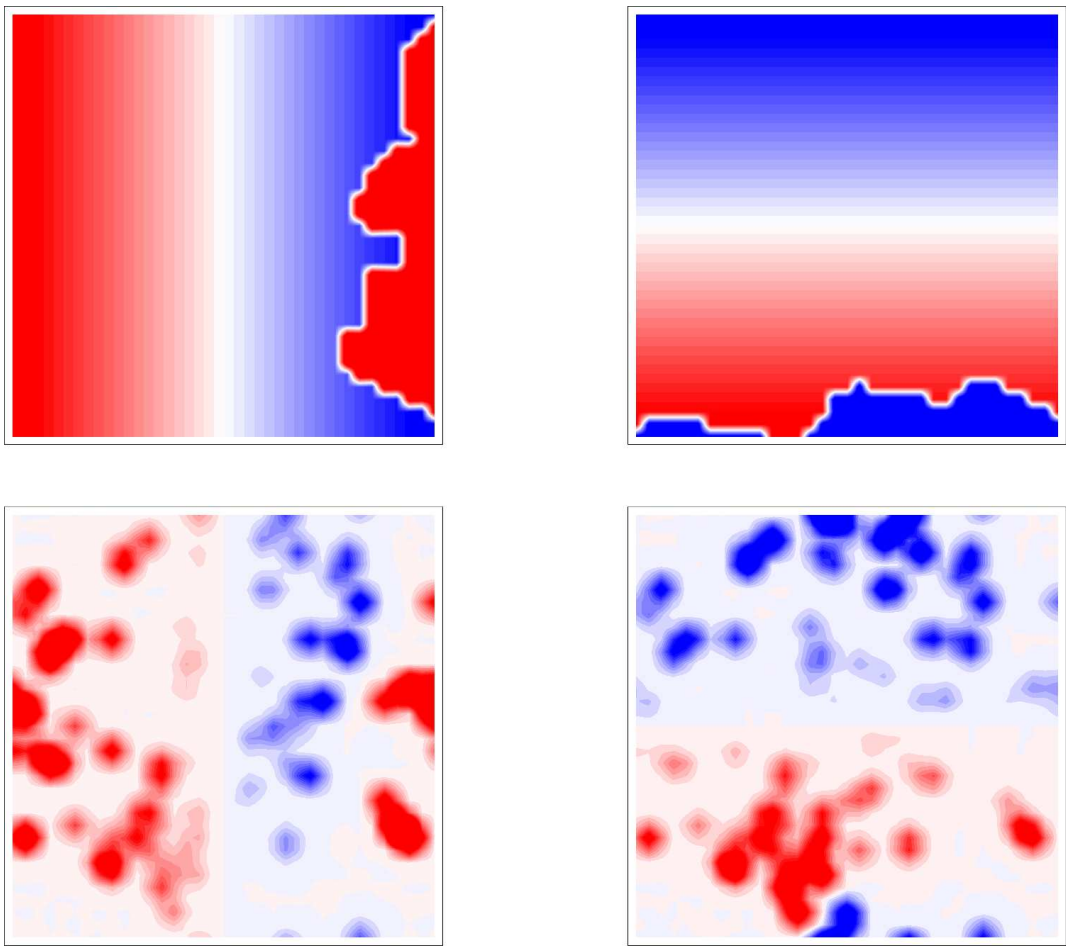}
\end{figure}

\begin{figure}
\caption{Contour plot of the function $\psi(c,\gamma)$ defined in equation (\ref{psidef}).}
\includegraphics{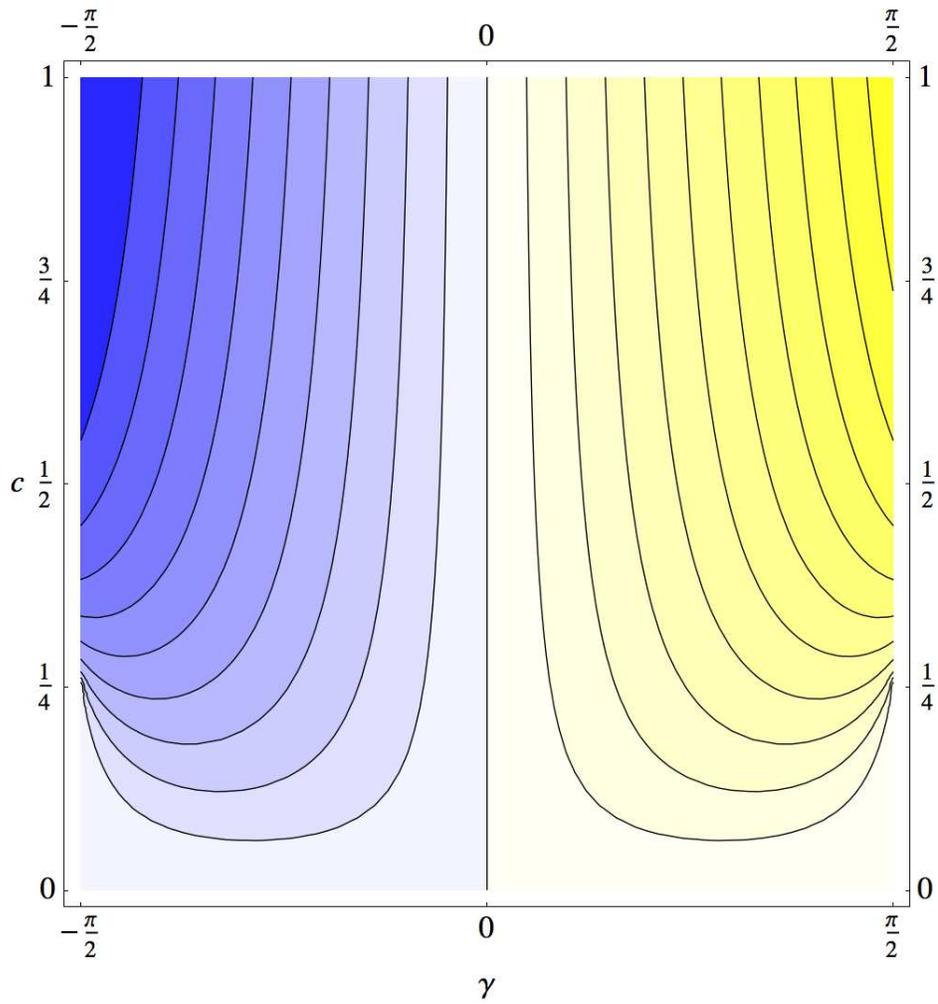}
\end{figure}

\begin{figure}
\caption{Example of an S-shaped particle whose unwrapping function has more than one discontinuity per crystal axis. Along the vertical edge of the unit cell (black square) we see that $u_y$ has three discontinuities (solid circles); $u_x$, along the horizontal edge, has only one discontinuity (open circle).}
\includegraphics{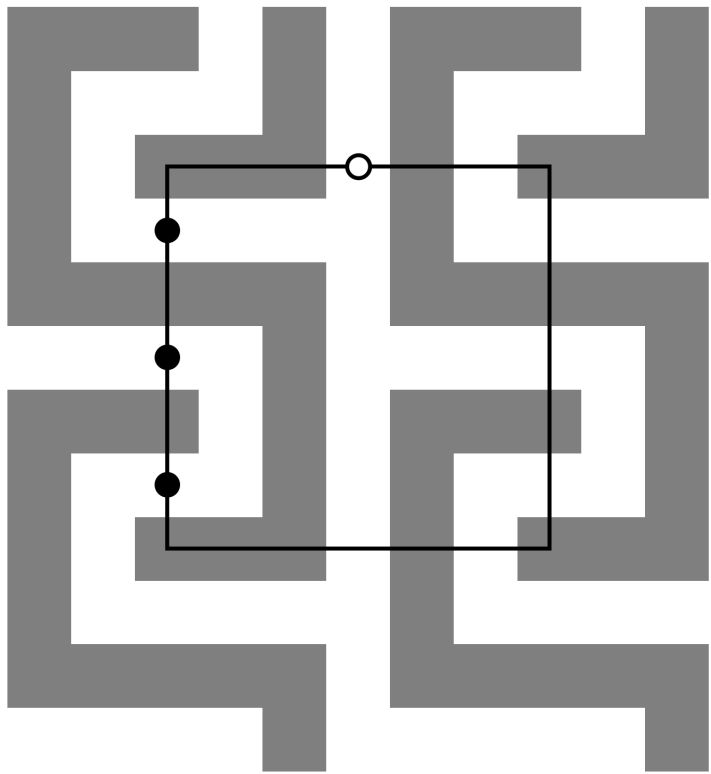}
\end{figure}

\begin{figure}
\caption{Low-pass filtered intensity data ($A$, on left) and intensity-gradient data ($\mathbf{B}$, on right) in the $h=0$ plane. Positive data are rendered yellow, negative values are blue.}
\includegraphics{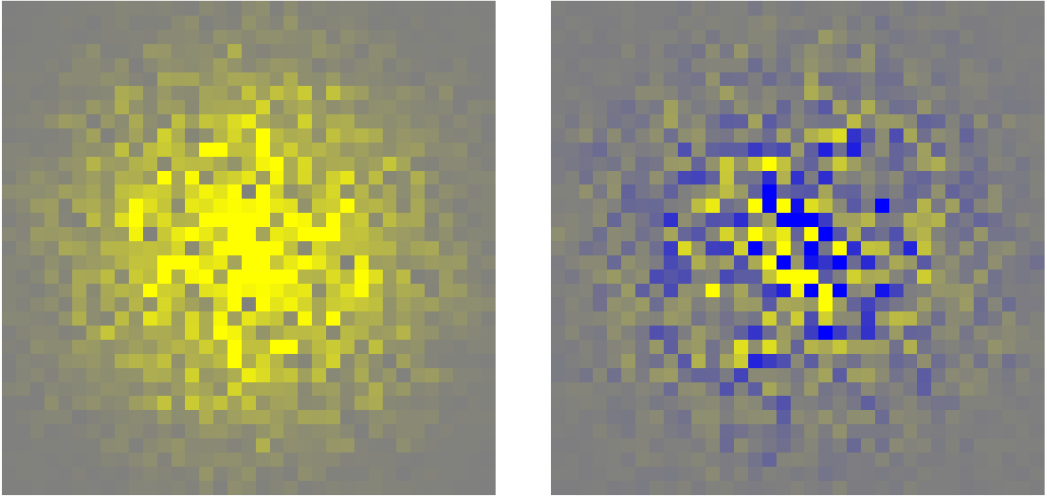}
\end{figure}

\begin{figure}
\caption{\textit{Top}: Electron density of lysozyme at the resolution used in this study. \textit{Bottom}: Reconstruction with $\mathrm{SNR} =2$ and known unwrapping function, rendered at slightly lower resolution.}
\includegraphics{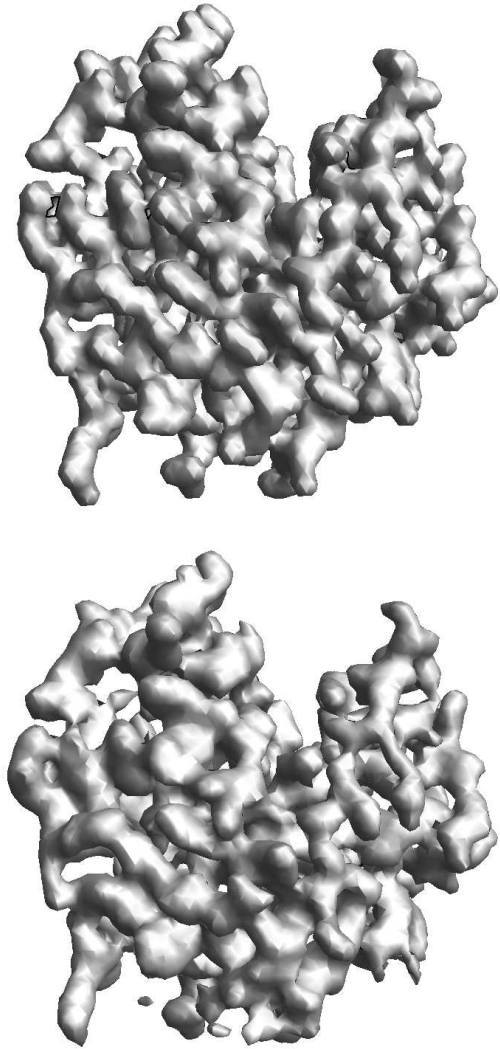}
\end{figure}

\begin{figure}
\caption{Update magnitudes ($\Delta$) for lysozyme reconstructions with known unwrapping function and $\mathrm{SNR}=20,10,5,2,1$ (highest SNR has lowest final $\Delta$).}
\includegraphics{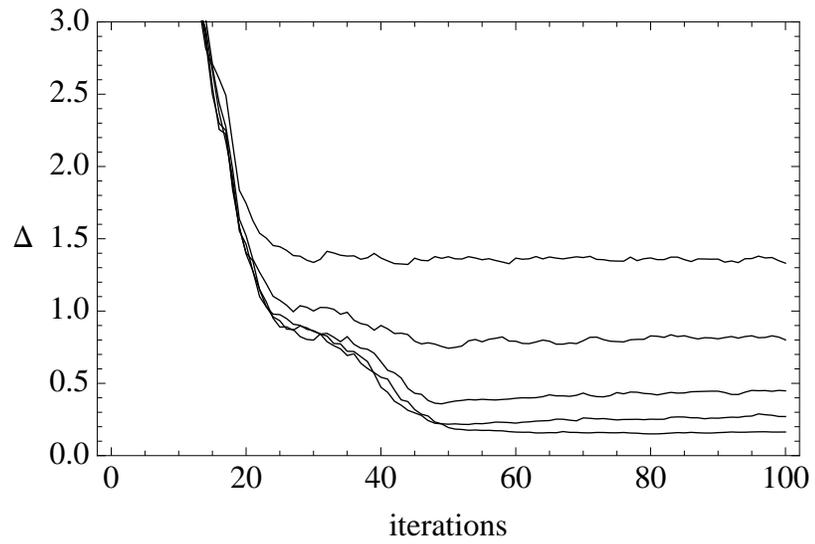}
\end{figure}

\begin{figure}
\caption{Evolution of the three wrapping surfaces (color contours) in a lysozyme reconstruction. Shown are the surfaces after update 5 (top row), 10, 20, 50 and 100 (bottom row). }
\includegraphics{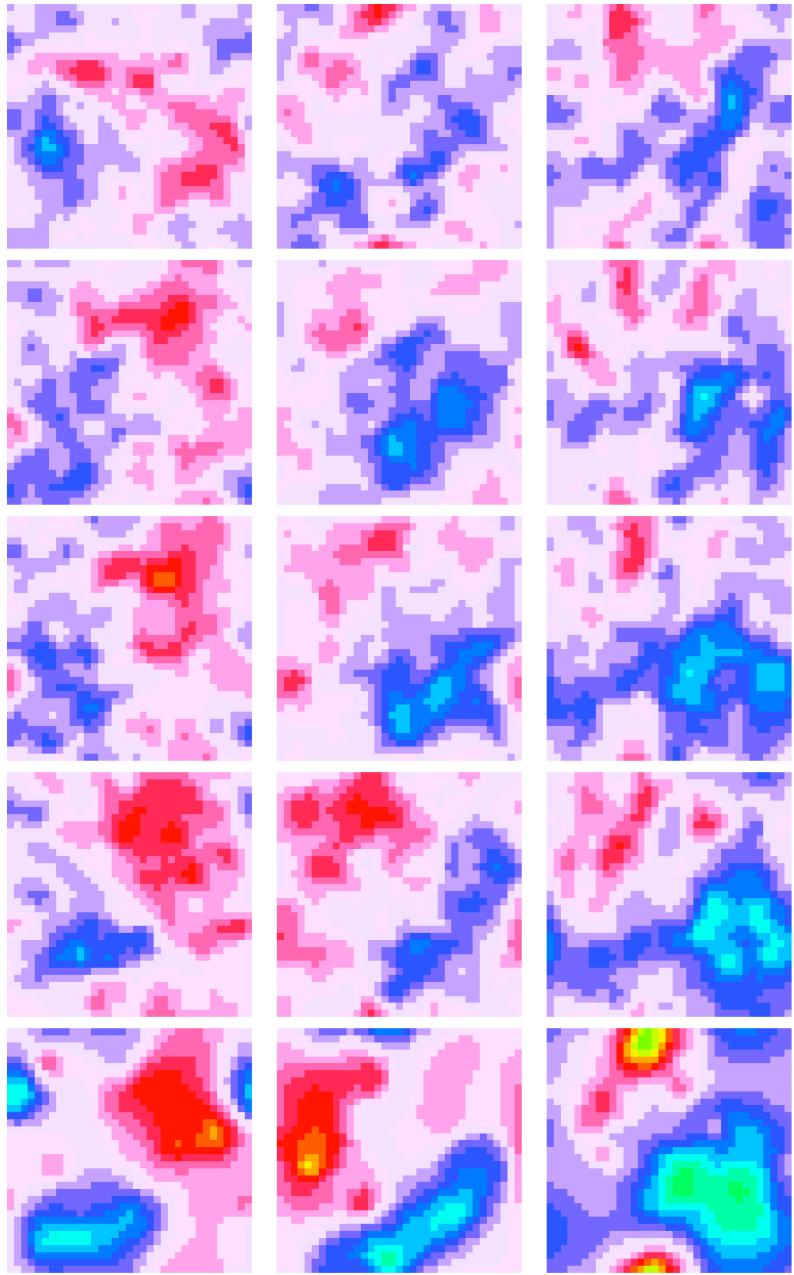}
\end{figure}

\begin{figure}
\caption{Update magnitudes ($\Delta$) for lysozyme reconstructions with unknown unwrapping function and $\mathrm{SNR}=20,10,5$. The regular peaks in $\Delta$ every 10 iterations reflect the update schedule of the wrapping function.}
\includegraphics{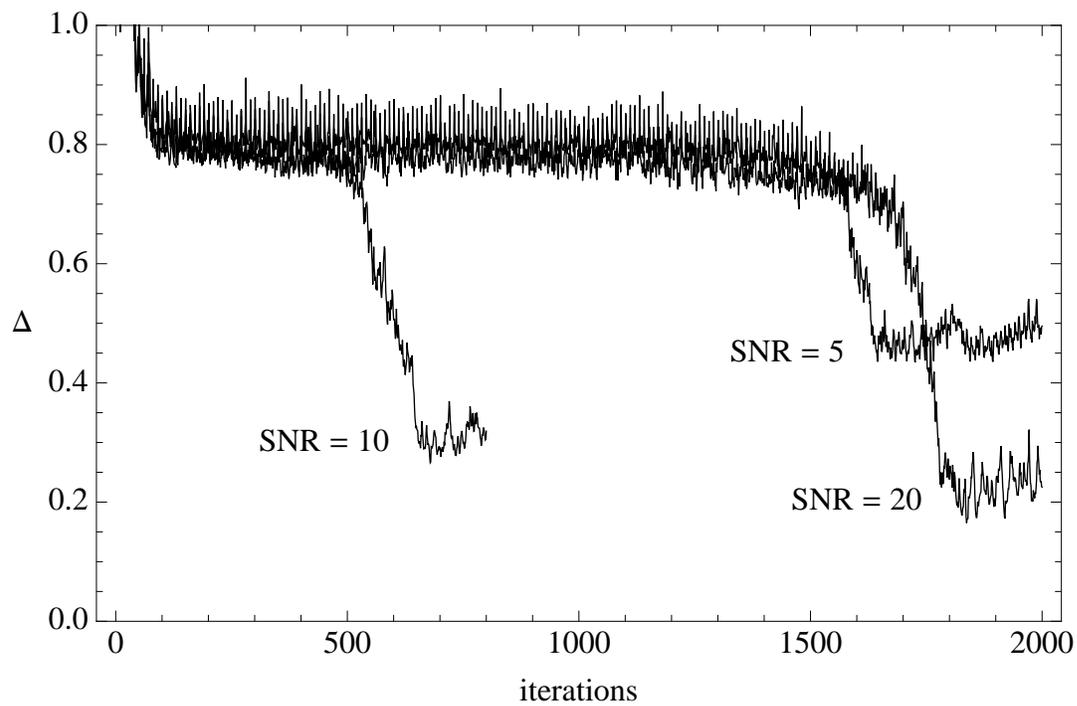}
\end{figure}

\begin{figure}
\caption{\textit{Top}: Electron density of lysozyme at reduced resolution. \textit{Bottom}: Reconstruction with $\mathrm{SNR} =5$ and unknown unwrapping function.}
\includegraphics{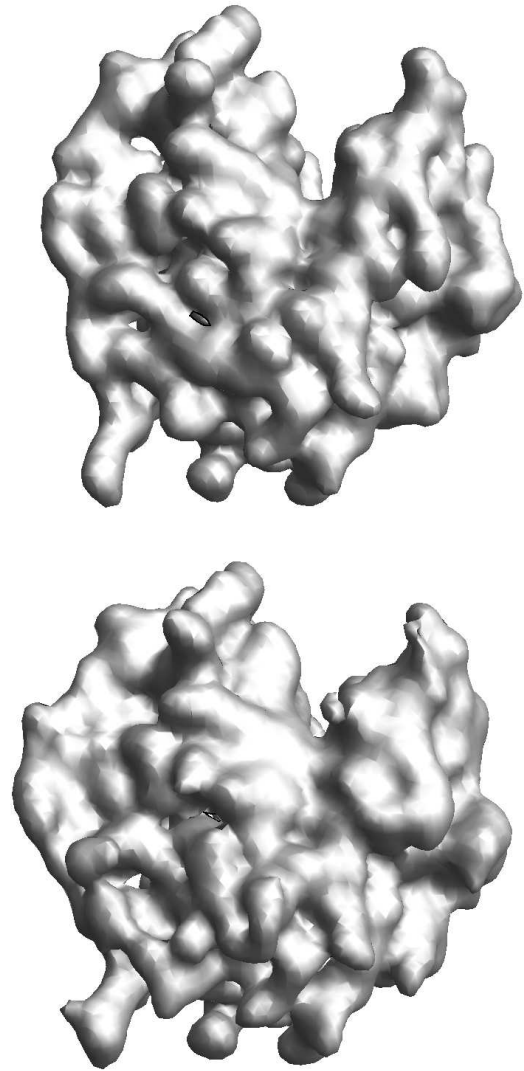}
\end{figure}

\begin{figure}
\caption{Cartoon of a crystal structure comprising two identical particles (triangles) per unit cell, but where the cell-occupancy of the particles varies along the surface of the nanocrystal.}
\includegraphics{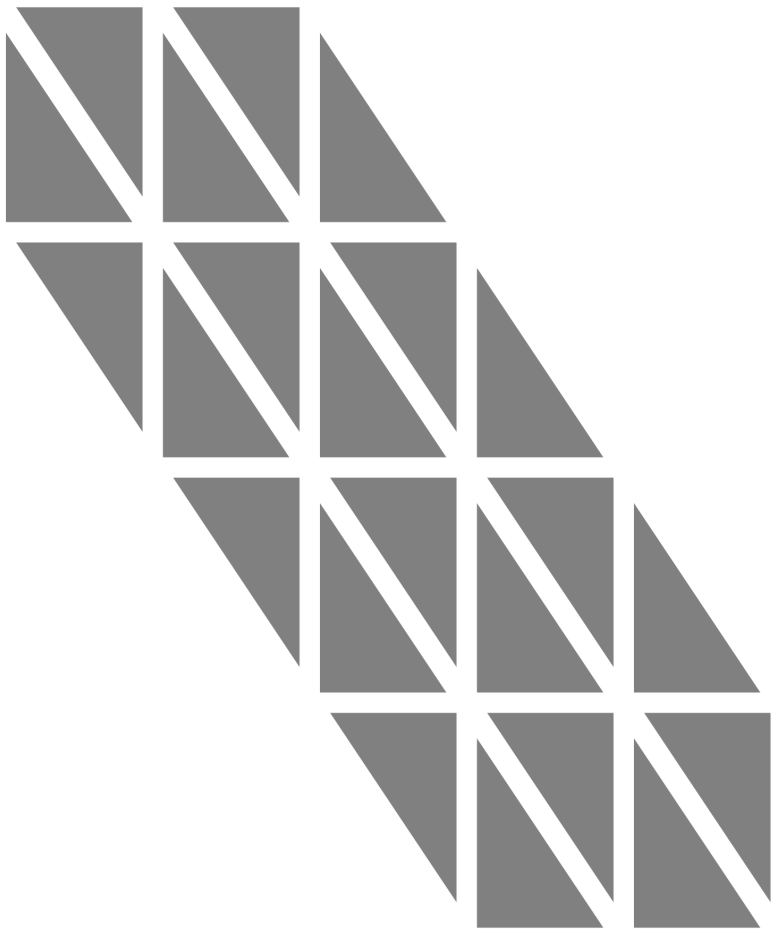}
\end{figure}

\end{document}